\documentclass[pre,aps,twocolumn,pacs]{revtex4}
\usepackage{epsfig}
\usepackage{bm}
\newcommand{\B}[1]{{\bm{#1}}}

\usepackage[latin1]{inputenc}
\newcommand{\beq}{\begin{equation}}
\newcommand{\eeq}{\end{equation}}
\newcommand{\bea}{\begin{eqnarray}}
\newcommand{\eea}{\end{eqnarray}}
\begin{document}
\title{Dissipative Visco-plastic Deformation in Dynamic Fracture:\\
 Tip Blunting and Velocity Selection}
\author{Eran Bouchbinder, Anna Pomyalov and Itamar Procaccia}
\affiliation{Dept. of Chemical Physics, The Weizmann Institute of
Science, Rehovot 76100, Israel}

\begin{abstract}
Dynamic fracture in a wide class of materials reveals ``fracture
energy" $\Gamma$ much larger than the expected nominal surface
energy due to the formation of two fresh surfaces. Moreover, the
fracture energy depends on the crack velocity,
$\Gamma\!=\!\Gamma(v)$. We show that a simple dynamical theory of
visco-plasticity coupled to asymptotic pure linear-elasticity
provides a possible explanation to the above phenomena. The theory
predicts tip blunting characterized by a dynamically determined
crack tip radius of curvature. In addition, we demonstrate
velocity selection for cracks in fixed-grip strip geometry
accompanied by the identification of $\Gamma$ and its velocity
dependence.
\end{abstract}
\pacs{PACS number(s):} \maketitle

{\bf Introduction}:  The dynamics of rapid fracture are still
posing a great challenge to theoretical physics \cite{FM99}. A
major difficulty in understanding the rich experimental
phenomenology is the coupling between very different scales; while
fracture is ultimately driven by the release of the large scales
elastic energy, the detailed deformation on the small scales near
the moving crack edge determines both the stress conditions and
energy dissipation that control the rate of growth and its
direction. Since the near crack edge deformation is extremely
complicated, it is common to lump all the near crack edge physics
into a phenomenological unknown fracture energy, leaving the
theory with a crucial missing ingredient. To better understand
these issues, consider a semi-infinite crack forming in an
infinite strip under fixed-grip boundary conditions. As the crack
proceeds at velocity $v$, a constant energy per unit advance per
unit time, $J$, is released from the elastic fields. In the
classical theory of Linear Elasticity Fracture Mechanics one
assumes that $J\!=\!v\Gamma$ where $\Gamma$ is the fracture
energy, putatively a constant, equal to the energy $2 \gamma_s$ of
forming two fresh surfaces. In fact, in a large variety of
materials, from steel to brittle plastics, the measured fracture
energy $\Gamma$ can be orders of magnitude larger than
$2\gamma_s$. Moreover, in many instances the measured fracture
energy depends on the velocity, $\Gamma\!=\!\Gamma(v)$. This is
important; cracks tend to reach a stationary velocity $v$ that
grows as a function of the loading and this is possible only if
$\Gamma$ depends on $v$. It is tempting to assert that some kind
of plastic processes may be responsible for this kind of findings.
Straightforward plasticity theory assumes that the stress fields
near the tip of a crack are tamed by plastic deformations and
remain at a value close to the yield stress $s_y$ of the material.
Alas, this value of stress is assumed to be smaller than the level
of stress necessary for overcoming the cohesive forces that bind
material together. Without a proper theory of the dynamical
processes occurring near the crack tip it remains very difficult
to explain all these observations in a coherent and
self-consistent manner.

{\bf Theoretical framework}: a promising formulation of
visco-plasticity was proposed recently by Falk and Langer
\cite{FL98}. As in all approaches to plasticity, one adds to the
usual  stress and elastic strain tensors $\sigma_{ij}$ and
$\epsilon^{el}_{ij}$,  a plastic component of the strain tensor,
denoted as a trace-less symmetric tensor $\epsilon^{pl}_{ij}$.
Essential to this theory  are the highly localized regions called
the ``shear transformation zones'' (STZ). These dominate the
plastic events and their presence is carried by two ``internal
state'' fields \cite{Lubliner}. One is a trace-less symmetric
tensor field $\B \Delta$ that is related to the orientations of
these zones and acts as a ``back stress''; the second is a scalar
field $\Lambda$ which stands for the density of these STZ. We
employ here the tensorial quasi-linear version of this model,
which is given by the following three coupled equations
\cite{FL00}:
\begin{eqnarray}
\dot{\epsilon}^{pl}_{ij} &=& \frac{1}{\tau}\left(\lambda  \Lambda
s_{ij} - \Delta_{ij}\right) \label{doteps}\\
\dot{\Delta}_{ij} &=&
\dot{\epsilon}^{pl}_{lk}\left(\delta_{il}\delta_{jk} -
\frac{s_{lk}}{2 \lambda s^2_y}\Delta_{ij}\right) \label{dotdel}\\
\dot{\Lambda} &=& \frac{s_{lk}\dot{\epsilon}^{pl}_{lk}}{2 \lambda
s^2_y} \left(1-\Lambda\right) \label{STZ} \ .
\end{eqnarray}
Here $s_{ij}$ is the 2D deviatoric stress tensor, $
s_{ij}\!\equiv\!\sigma_{ij}\!-\!\frac{1}{2}$Tr$~\B\!\sigma
\delta_{ij}$, $\tau$ is a typical time scale and $\lambda$ is a
parameter of dimension inverse stress that measures the plastic
strain rate sensitivity to stress. This set of equations and
further developments were studied in considerable detail in other
contexts \cite{STZ_ref,Langer00}. In this Letter we study the
implications of this model for dynamic fracture. The reader should
note that in principle the time scale $\tau$ may depend strongly
on $s_{ij}$, maybe even diverging for $|s_{ij}|\!\ll\! s_y$. With
such divergence one can connect smoothly to elastic behavior far
away from the crack tip. In our analysis below we take $\tau$ to
be constant, but on the other hand allow the density $\Lambda$ of
STZ to decay to its pre-deformation small value far away from the
crack tip. In this we differ from previous treatments
\cite{Langer00} where $\Lambda$ was taken to be unity. We will
show below that the dynamics of $\Lambda$ are interesting and
important, in addition to allowing us a smooth coupling to
elasticity.

Consider then a semi-infinite crack in a 2D infinite strip of a
given width, loaded in fixed-grip boundary conditions such that
the energy release rate $G\!\equiv\!J/v$ is experimentally
controlled \cite{Freund}. The crack advances steadily at a
velocity $v$ (to be found self-consistently). The effect of the
plastic response should be strongest near the tip of the crack,
while far away from the tip linear elasticity can be safely
employed. Like in \cite{Langer00} we expect the crack tip radius
of curvature $R$ to adjust itself dynamically, allowing the crack
tip to blunt. The degree of blunting will be connected
self-consistently to the steady velocity $v$ \cite{BS}. While we
adopt below a number of approximations, we insist on the
conservation of energy in the sense that the stored elastic energy
is mainly dissipated near the blunted tip (a feature that appears
missing in \cite{Langer00}), though part of it can be locked in
the crack wake. Sensitive to the experimental finding that
$\Gamma\!\gg\!2\gamma_s$, we will assert that the dominant
contribution to $\Gamma$ comes from the plastic work which we
assume to include both the energy dissipated and the residual
energy locked in the wake \cite{Freund}; this assumption
translates to
\begin{equation}
\Gamma(v, R) \simeq \frac{1}{v}\int_A s_{ij}
\dot{\epsilon}^{pl}_{ij} dA = G \ , \label{bal}
\end{equation}
where $A$ is the area in which the plastic dissipation is not
vanishingly small. This equation should be considered together
with the kinematic relation at the crack tip ($\theta=0$),
\begin{equation}
v \simeq R~\dot{\epsilon}^{pl}_{\theta \theta}(x\!-\!vt\!=\!R,0)
\label{kinematic} \ ,
\end{equation}
where we have assumed that the plastic strain rate dominates the
elastic one.

In order to solve Eqs. (\ref{doteps})-(\ref{kinematic}) we need an
explicit solution for $s_{ij}$. The solution for the deviatoric
stress is not independent of the visco-plastic constitutive model
in Eqs. (\ref{doteps})-(\ref{STZ}). The coupling is via the
relation of the deviatoric stress to the elastic part of the
strain tensor, which needs to add up to the plastic part such that
the total strain satisfies the usual compatibility relation
\cite{Lubliner}. We thus expect $s_{ij}$ to deviate from the usual
linear-elastic behavior. To quantify this deviation we introduce
the functions $h_{ij}(R/r, \theta,v)$ (no summation convention)
\begin{equation}
s_{ij}(x\!-\!vt,y,v) = \frac{K_{_{\rm I}}(v)
\hat{\Sigma}_{ij}(\theta,v)}{\sqrt{2\pi r}} h_{ij}(R/r, \theta,v)
\ , \label{sij}
\end{equation}
where $r\!\equiv\!\sqrt{(x\!-\!vt)^2+y^2}$,
$\theta\!=\!\tan^{-1}\left(y/(x\!-\!vt)\right)$ are polar
coordinate system whose origin is at distance $R$ behind the tip
and $\hat{\Sigma}_{ij}$ are known universal functions
\cite{Freund}. The unknown functions $h_{ij}$ incorporate both a
dynamic effect ($s_{ij}$ appears now also in the time derivative
of the compatibility relation) and a geometric effect (tip
blunting) on the stress. Clearly, these functions approach unity
as $R/r\!\to\!0$ and the remaining part is the sharp-tip
linear-elastic asymptotic solution with the stress intensity
factor $K_I(v)$ \cite{Freund}
\begin{equation}
\label{SIF}
K_{_{\rm I}}(v) = \sqrt{\frac{E~G}{A_{_{\rm I}}(v/c_s,
\nu)}} \  .
\end{equation}
Here $A_{_{\rm I}}$ is a known universal function \cite{Freund},
and $c_s$, E and $\nu$ are the shear wave speed, Young's modulus
and Poisson's ratio respectively. Below we approximate the
functions $h_{ij}$ in order to integrate Eqs.
(\ref{doteps})-(\ref{STZ}) and  find self-consistently $v$ and $R$
according to Eqs. (\ref{bal}) and (\ref{kinematic}).

Before proceeding to our results, we recall a different
line of investigation taken in \cite{FH85}. There the near
{\em sharp} tip stress distribution of Eq. (\ref{sij}), with
$h_{ij}\!=\!1$, was used to compute the plastic strain rate using
a phenomenological visco-plastic constitutive law. Then the
dissipation rate associated with the plastic strain rate was
computed and added to an unknown velocity-independent dissipation
to be equated to the energy release rate $G$. The two approaches,
ours and the one presented in \cite{FH85}, are approximate in
nature and even share some of the underlying assumptions, but in
fact are very different. Indeed, we use elastic stress
solutions inside the plastic zone as in \cite{FH85} and also
demand explicitly energy conservation, but the similarity ends here.
We consider the sharp tip assumption unjustified and
by asserting that the crack tip advances mainly by plastic flow,
arrive at Eq. (\ref{kinematic}) that has no analog in \cite{FH85}; as a result
the unrealistic stress singularity is cut off. Moreover, in contrast to \cite{FH85} we assert that
the plastic dissipation is dominant. Finally, we incorporate a {\em dynamical rate-and-state}
formulation of visco-plasticity that is fundamentally different
from the phenomenological constitutive law employed in
\cite{FH85}.

{\bf Analysis}: To analyze the various regimes of solutions we
non-dimensionalize the equations by measuring stress in units of
$s_y$, time in units of $\tau$ and velocity in units of $c_s$.
Having only one typical time scale, we expect that
\begin{equation}
\frac{v\tau}{R} \sim {\cal O}(1) \ . \label{time}
\end{equation}
In light of Eq. (\ref{kinematic}), sufficiently near the tip the
dimensionless plastic strain rate is also of ${\cal O}(1)$. In seeking solutions
we keep this physical constraint in mind.
\begin{itemize}
\item{First regime of solutions, apparently not physical}
\end{itemize}
An analytically tractable parameter regime is $\lambda s_y\!\ll\!1$. Observe
Eq. (\ref{doteps}), and note that the LHS is of ${\cal O}(1)$. The
fact that $\lambda s_y\!\ll\!1$ renders $s_{ij}/s_y\!\gg\!1$,
sending $\Delta_{ij}$ to be $\ll\!1$, according to  Eq.
(\ref{dotdel}). It also follows that $\Lambda\!\to\!1$ very
rapidly. In this limit the set of equations
(\ref{doteps})-(\ref{STZ}) simplifies in the region of interest
near the tip to
\begin{equation}
\dot \epsilon^{pl}_{ij}\simeq \frac{\lambda}{\tau} s_{ij} \ .
\label{limit1}
\end{equation}
Therefore the equations for the components of $\dot
\epsilon^{pl}_{ij}$ decouple in this limit, identifying $\lambda$,
as expected, with the visco-plastic analog of an inverse shear
modulus. To proceed, we assert that the predominant $v$-dependence
in Eq. (\ref{sij}) is carried by $K_{_{\rm I}}(v)$. This is
justified since $\hat{\Sigma}_{ij}$ are qualitatively the same for
velocities below the Yoffe threshold \cite{Freund} and the
$v$-dependence in $h_{ij}$ can be neglected according to
\cite{Freund, FH85}. We can substitute now Eq. (\ref{limit1}) in
Eqs. (\ref{bal}-\ref{kinematic}) and use the approximate $s_{ij}$
of (\ref{sij}) we end up with
\begin{eqnarray}
v&\sim& \frac{\lambda G}{\tau} \ . \label{v}\\
R &\sim& \frac{\tau v}{\lambda E} ~A_I(v/c_s,\nu) \sim \frac{G}{E}~
A_I\left(v(G)/c_s,\nu\right) \ . \label{R}
\end{eqnarray}

Note that this solution has interesting aspects: we find a
self-consistent steady velocity that increases linearly upon
increasing $G$. This implies that $\Gamma(v)\!\sim\!v$. This
solution is realized self-consistently by having also $R$
increasing with G. Since $A_I\!\to\!1$ with zero slope when
$v\!\to\!0^+$ \cite{Freund}, $R$ starts linearly with $G$ and
later turns nonlinear according to the dependence of $A_I$ on $G$.
Unfortunately, this solution allows very large stress
concentrations near the tip, i.e. $s_{ij}/s_y\!\gg\!1$ and as we
are unaware of experimental observations indicating such high
stress levels, we propose that the limit $\lambda s_y\!\ll\!1$ is
not physically relevant.
\begin{itemize}
\item Second regime of solutions, apparently physical
\end{itemize}
The range of parameters that furnishes an interesting solution is
$\lambda s_y\!\gtrsim\!1$. In this range the constitutive
equations (\ref{doteps})-(\ref{STZ}) are fully coupled and the
``internal state'' fields ${\bf \Delta}$ and $\Lambda$ play an
important role in the dynamics. To see this we first note that in
order to satisfy Eq. (\ref{time}) in the near tip region the
combination $\lambda \Lambda s_{ij}$ should be of ${\cal O}(1)$.
According to Eqs. (\ref{doteps}) and (\ref{STZ}) this condition
can be realized with $\Lambda\!<\!1$ and $s_{ij}\!\gtrsim\!s_y$.
In that case, according to Eq. (\ref{dotdel}), the ``back stress''
$\Delta_{ij}$ is comparable to $\lambda \Lambda s_{ij}$ and indeed
resists plastic deformations. Thus for $\lambda s_y\!\gtrsim\!1$
we expect a solution with $s_{ij}\!\gtrsim\!s_y$, where $\Lambda$
does not saturate rapidly and the dynamics of $\B \Delta$ are
important. To shed more light on the nature of this solution and
to substantiate the first solution, we consider the equations
numerically.
%%%%%%%% FIGURE 1 %%%%%%%%%%%%%%%%%%%%
\begin{figure}
\centering \epsfig{width=.35\textwidth,file=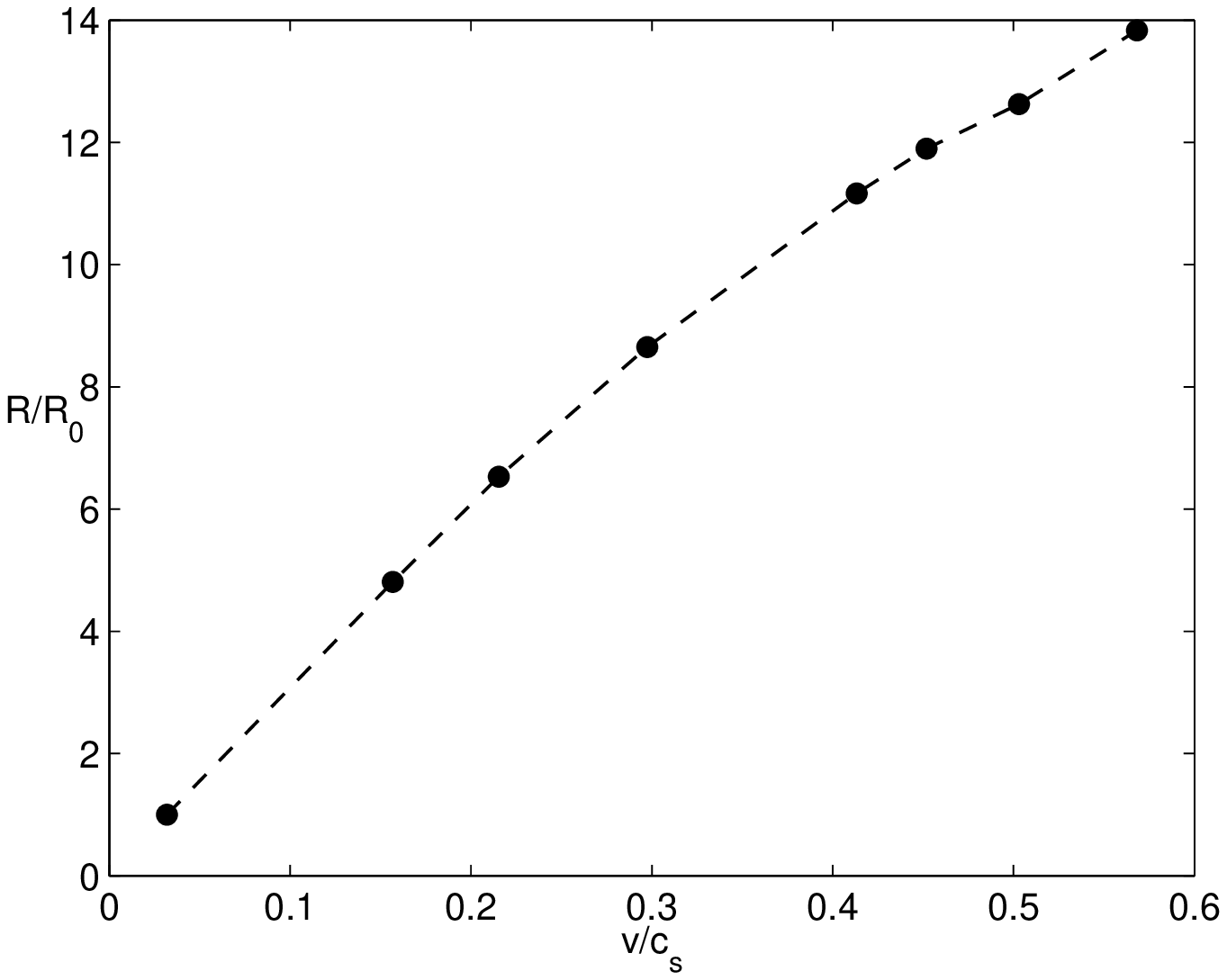}
\centering \epsfig{width=.35\textwidth,file=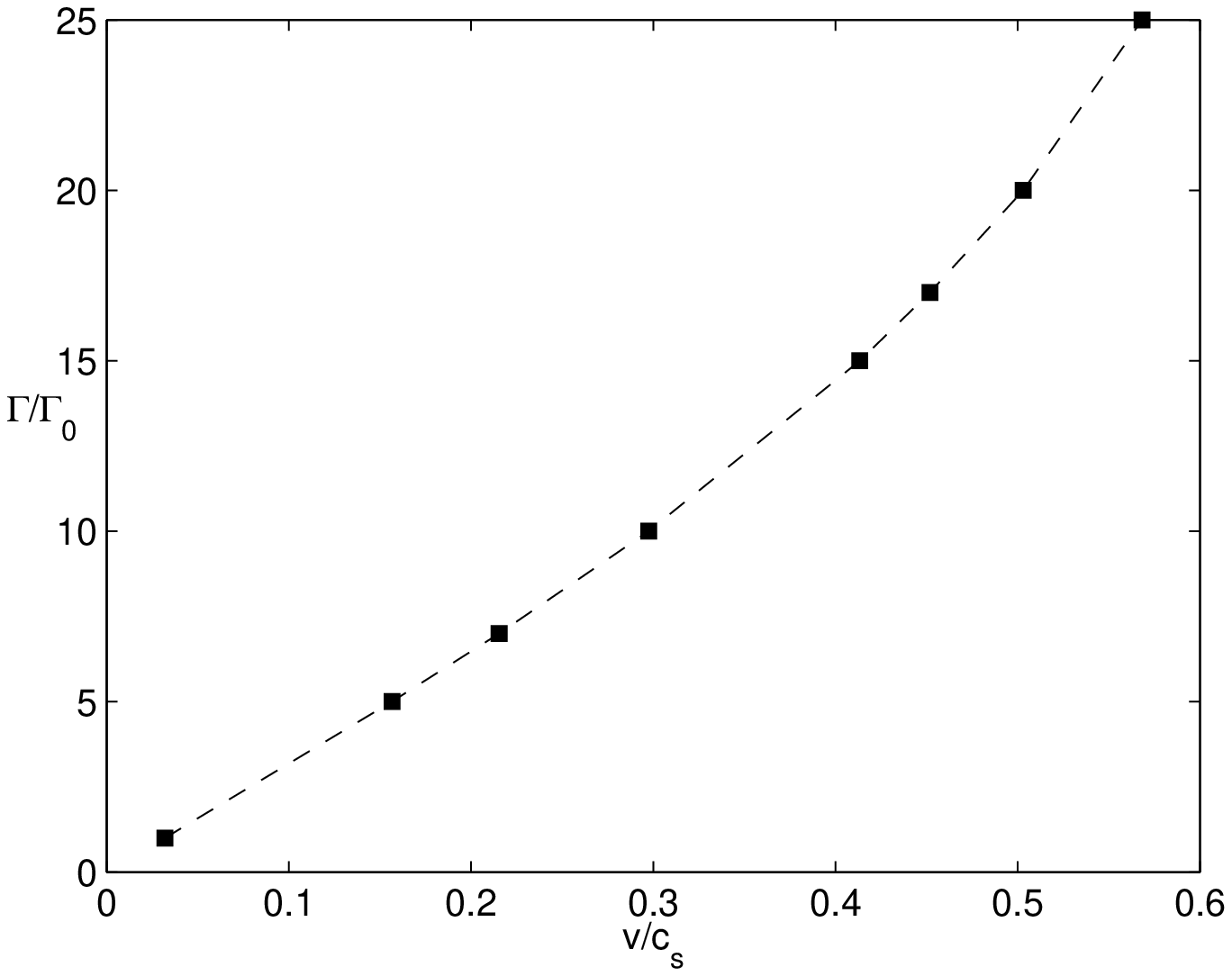}
\caption{Solutions for $\lambda s_y\!=\!1$. Upper panel: The tip
radius $R$ normalized by the first data point as a function of the
normalized  velocity $v/c_s$. Lower panel: The fracture energy
$\Gamma$ normalized by the first data point as a function of the
normalized  velocity $v/c_s$.} \label{results}
\end{figure}
%%%%%%%%%%%%%%%%%%%%%%%%%%%%%%%%%%%%%%

{\bf Numerics}: For the sake of numerical solutions we approximate
the solution for $s_{ij}$ by allowing the visco-plastic processes
to blunt the tip and then solve approximately for the stress field
up to the blunted tip via quasi-static linear elasticity. This
approach inevitably violates compatibility, but is expected to
yield reasonable approximations as shown in \cite{Langer00}. Thus
the main effect of the visco-plastic processes on the stress is
via the introduction of the length scale $R$ near the tip, hence
neglecting all the $v$-dependence except for $K_{_{\rm I}}(v)$, as
was explained above. In that case, outside the crack we solve for
the stress field with boundary conditions
\begin{equation}
\sigma_{rr}(r\!\!=\!\!R,\theta)=\sigma_{r \theta}(r\!\!=\!\!R,\theta)=0\
, \label{bc}
\end{equation}
for a finite range of angles (say,
$-\pi/3\!\!<\!\!\theta\!\!<\!\!\pi/3$). Since
$r$ cannot go to zero, we can add to the usual $r^{-1/2}$ solution
more ``divergent" powers; explicitly, we add terms to $\sigma_{ij}$
of the form
\begin{equation}
\frac{K_{_{\rm I}}(v)}{\sqrt{2 \pi r}} \frac{R}{r}
f_{ij}(\theta) \ , \quad
\frac{K_{_{\rm I}}(v)}{\sqrt{2 \pi r}}
\left(\frac{R}{r}\right)^2 g_{ij}(\theta) \label{newterms} \ .
\end{equation}
The pre-factors of the new terms can be selected such as to best
approximate Eq. (\ref{bc}) in the sense that relative to the
non-vanishing component $\sigma_{\theta\theta}$ the ratios
$|\sigma_{rr}/\sigma_{\theta\theta}|$ and $|\sigma_{r
\theta}/\sigma_{\theta\theta}|$ are bounded below 1\%. This
solution is naturally matched to the usual linear-elastic
asymptotic solution far from the tip, while capturing
qualitatively the elimination of the common sharp tip singularity
near the tip.

At this point we solve Eqs. (\ref{doteps})-(\ref{STZ}) along the
line $\theta=0$. The calculation of the integrals in Eq.
(\ref{bal}) should be implemented with care. In contrast to
classical theories of plasticity, here there is no sharp boundary
between the visco-plastic and elastic solutions: the deviatoric
stress $s_{ij}$ and therefore $\dot{\epsilon}^{pl}_{ij}$ decay
{\em algebraically}. Thus we should choose a reasonable cutoff for
which the integral in our model converges. For that purpose, we
compute the integrand of (\ref{bal}) along the line $\theta=0$ and
in particular determine the position $x=W$ at which the plastic
work rate $s_{ij}\dot{\epsilon}^{pl}_{ij}$ reaches 2\% of its
maximal value at the tip. Using this upper limit of integration
the value of the integral becomes insensitive to the long tail of
the plastic rate of work. We then estimate the 2D integral as the
1D integral times $\alpha W$, with $\alpha$ being a dimensionless
factor representing the full contribution from the loading {\em
and} unloading regions (where in the latter the model is
inapplicable \cite{STZ_ref}). The model equations exhibit a
solution when $\alpha$ is chosen in the range $0.17-0.25$.
Finally, the 1D integral can be computed over time rather than
space (since all the functions depend on $x\!-\!vt$). This allows
us to integrate Eqs. (\ref{STZ}) in time, capturing the essential
history dependence of $\epsilon^{pl}_{ij}$ on the internal state
fields $\Lambda$ and ${\bf \Delta}$. At this point, Eqs.
(\ref{bal})-(\ref{kinematic}) can be regarded as defining two
surfaces in an abstract $v\!-\!R$ space. We  search graphically
for the zero of the intersection line of these surfaces as
parametrized by $G$.

Following this procedure we solved for $R$ and $v$ for the two
ranges of $\lambda s_y$ considered above. First, we analyzed the
model with $\lambda s_y=0.01$, $\nu=0.3$, $E/s_y=50$,
$\alpha=0.25$,
$\epsilon^{pl}_{ij}(t\!=\!0)=\Delta_{ij}(t\!=\!0)=0$, and
$\Lambda(t\!=\!0)=0.01$ and compared the numerical solution to the
analytic counterpart. We verified that for similar small initial
values of $\Lambda$ the results are rather insensitive to the
exact value. The numerical solution matches Eqs.
(\ref{v})-(\ref{R}) extremely well. Nevertheless, the
dimensionless combination $G/s_y R$ in the numerical solution
equals $\approx 3\times 10^3$. Taking PMMA \cite{PMMA} as a
representative for the relevant class of materials, this number
implies $R \approx 5 nm$. This length scale is much smaller than
expected for PMMA, thus supporting our proposition that the range
$\lambda s_y \ll 1$ is not physically relevant. On the other hand,
in the parameter range $\lambda s_y\!\gtrsim\!1$ we verified that
both ``internal state'' fields $\B \Delta$ and $\Lambda$ play an
important role in the dynamics and the dependence of both $v$ and
$R$ on $G$ is non-linear. The dependence of $R$ and $\Gamma$ on
$v$ for $\lambda s_y\!=\!1$ is shown in Fig. \ref{results}.
Numerically we have found that to a very good approximation
\begin{equation}
\Gamma \propto v A_I(v)\ , \quad R\propto v/A_I(v)\ . \label{approx}
\end{equation}
Since $A_I$ appears in the theory always in the combination $A_I/E$ (cf. Eq. (\ref{SIF})),
we can write $\Gamma \propto v s_y A_I(v)/E$,  $R\propto vE/s_yA_I(v)$ where
we non-dimensionalized $E$ by $s_y$. Rearranging these equations and
keeping dimensions right we end up with the implicit relations
\begin{equation}
v = \frac{ E G}{\tau s^2_y A_I(v)}f_v(\lambda s_y)\ , \quad R=\frac{E^2 G}{s_y^3A^2_I(v)}
f_R(\lambda s_y)\ ,\label{final}
\end{equation}
where $f_v$ and $f_R$ are two dimensionless nonlinear functions
computed from the numerical solution, reflecting the importance of
the internal state variables $\Lambda$ and $\B \Delta$ in the
solution. Note that the solution for $R$ in (\ref{final}) differs
from the one provided in \cite{Langer00} even in the quasistatic
limit, reflecting the difference between the two approaches,
mainly in the use here of Eq. (\ref{bal}). In this equation $G$
appears twice, once on the RHS and once through the stress
intensity factor on the LHS, thus changing the way that $E$
appears in the final solution.

Finally, the dimensionless combination $G/s_y R$ for this class of
solutions is typically of $ {\cal O}(1)$. Applying the analysis to
PMMA we estimate $R\approx 15 \mu m$, which is an acceptable
estimate for this class of materials. We propose therefore that
the solutions in the range $\lambda s_y\!\gtrsim\!1$ are
physically relevant for materials in which the main dissipation
mechanism is visco-plasticity. The stresses transmitted to the
crack tip itself are moderately larger than $s_y$, supposedly
sufficient to overcome the cohesive forces binding the material
together, and the plastic zones are relatively thin and changing
smoothly towards the linear elastic asymptotic fields.

{\bf Summary}: We have employed a theoretical framework of
visco-plasticity to propose a direct measure of the fracture
energy in dynamic fracture for a class of materials where the
fracture energy exceeds significantly the bare surface energy. The
fracture energy is a function of the crack velocity, rationalizing
the experimental observations of velocity selection for cracks in
a fixed-grip strip geometry. This selection is accompanied by the
blunting of the tip, with a radius that is self-consistent with
the velocity. The theory, within the stated approximations,
predicts that both $\Gamma$ and $R$ are monotonically increasing
functions of the velocity $v$ in the considered range. In the
physical range of parameters ($\lambda s_y\!\gtrsim\!1$) the
stress at the tip exceeds the yield-stress $s_y$, affording
rupture of the cohesive bonds. In this range of parameters the
plastic zone is relatively thin with respect to the tip radius.

The STZ model as used here cannot be considered a final theory of
visco-plasticity \cite{STZ_ref}. A rigorous coupling to elasticity
and a consistent description of the wake of the crack are still
missing here. Nevertheless we believe that the encouraging
approximate results presented here are generic to any reasonable
model that addresses the complex state of deformation near a
moving crack tip. In future work one should invest effort in
improved models where the approach can be tested in some
quantitative detail. In such future work the questions of crack
tip instabilities that were not addressed here at all should be
directly addressed and explained.

{\bf Acknowledgments}: We thank E. Brener, R. Spatschek and J. S.
Langer for very useful discussions. E.B. is supported by the
Horowitz Complexity Science Foundation. This work had been supported
in part by the Minerva Foundation, Munich, Germany, and the Israel Science
Foundation.


\begin{thebibliography}{99}
\bibitem{PMMA}
The material parameters used are $E\!=\!2.3GPa$, $s_y\!=\!60M\!Pa$
and $G\!\simeq\!10^3 J/m^2$.
\end{thebibliography}
\end{document}